\documentclass[epj,final]{svjour}[13.07.2000]

\newcommand{\ba}{\begin{eqnarray}}
\newcommand{\ea}{\end{eqnarray}}
\newcommand{\baa}{\begin{array}}
\newcommand{\eaa}{\end{array}}

\newcommand{\beq}{\begin{equation}}
\newcommand{\eeq}{\end{equation}}

\newcommand{\Cg}{{\mathcal C}_g}

\newcommand{\tr}{\mathrm{tr}}

\newcommand{\comment}[1]{}

\begin{document}

\title{ Scalar Mesons in a Chiral Quark Model with Glueball.}
\author{D. Ebert\inst{1}, M. Nagy\inst{2},
M. K. Volkov\inst{3}, \and V. L. Yudichev\inst{3}}
\institute{ Institut f\"ur Physik, Humboldt-Universit\"at zu Berlin, Invalidenstrasse 110, D-10115 Berlin, Germany
\and Institute of Physics, Slovak Academy of Sciences, 842 28 Bratislava, Slovakia
\and Bogoliubov Laboratory of Theoretical Physics,
Joint Institute for Nuclear Research,141980 Dubna, Russia}
\mail{volkov@thsun1.jinr.ru}

\abstract{
Ground-state scalar isoscalar mesons
and a scalar glueball are described in a $U(3)\times U(3)$
 chiral quark model of the Nambu--Jona-Lasinio
(NJL) type with  't Hooft interaction.
The latter interaction produces singlet-octet mixing in
the scalar and pseudoscalar sectors.
The glueball is introduced
into the effective meson Lagrangian as a dilaton
 on the base of scale invariance.
The mixing of the glueball with scalar isoscalar quarkonia
and amplitudes of their decays into two pseudoscalar mesons are
shown to be proportional to current quark masses,
vanishing in the chiral limit.
Mass spectra of the scalar mesons and the
glueball and their main modes of strong decay are described.
}

\maketitle


\section{ Introduction}

Scalar mesons play an important role in the strong interactions of
elementary particles~\cite{X1,Volk_86}.
They are also significant for a correct
description of nuclear interactions~\cite{X2}.
Today, 19 scalar
resonances are observed in the energy interval from 0.4 to 1.7 GeV
\cite{PDG}.
However, their nature is not yet quite clear.
Are they two-quark or four-quark states?
Which of them belong to the ground state nonet of the $U(3)$
 flavour group; and which, to the excited one?
Is there a glueball among them, and where does it lie?
These are intriguing questions, where to find answers,  many attempts
are currently being made by different authors.
From a quick survey of what has been done by now,
one could conclude that we do not yet approach
the solution.
A lot of models,
  (see, e.g., \cite{gb:models1,gb:models2,gb:models3,gb:models4,gb:models5,gb:models6,gb:models7,gb:models8,gb:models9,gb:models10,Shakin,Shakin:2000,ECHAJA1999,Amsler,Naris_98,Aniso_98,Tornqvist,Jami_96})
has been suggested, but none of the
approaches has given us the clue that could allow us to clarify the true nature of
scalar mesons once for all.  Moreover, one might be
disappointed by the fact that
different models give different results that do not much overlap,
being in some cases controversial.

In our work, we describe
scalar isoscalar mesons and their mixing with scalar glueballs.
As we learn from QCD, if there were no quarks in the world,
gluons themselves could form bound objects due to a large coupling strength of
gluon self-interactions. This stimulated the search for
bound gluon systems both in experiment and in theory.
As the gluon carries no flavour, lepton or baryon quantum number,
glueballs must be searched among isoscalar mesons.
Indeed, the simplest gluonic formation possesses the quantum numbers of
a scalar isoscalar meson.
If one looks into
tables of experimental data, one finds a large number
of scalar isoscalar states that can be mixed states of
quarkonia, multi-quark systems, hybrids, and glueballs.
To distinguish a glueball is really a difficult task,
because we have no reliable test that would give us the truth.

As the perturbative approach does not work here,
different phenomenological models and lattice simulations
are involved in the study.
From recent results
\cite{Sexton,LeeWeingarten,VaccarinoWeingarten}
one can conclude
that it is most probably that glueballs are  real objects of our world.
There exist numerical estimates for probable masses of glueballs,
however still in the world without quarks. Lattice calculations
report that the lightest scalar glueball should be found between
1.5 and 1.7 GeV.

Amsler \cite{Amsler}
considered
the state $f_0(1500)$ as a candidate for the scalar glueball.
QCD sum rules \cite{Naris_98} and the K-matrix method \cite{Aniso_98}
showed that both $f_0(1500)$ and $f_0(1710)$ are mixed states with
large admixture of the glueball component. Moreover, QCD sum rules \cite{Naris_98}
require that light glueballs (below 1 GeV) should exist, which is in
contradiction with what lattice calculations suggest.

A glueball cannot be searched  without investigating
the nature of the rest of scalar mesons that are not heavier than,
2 GeV and which we consider mostly as formed by quark-antiquark pairs.
All the bound isoscalar $q\bar q$ states are allowed to mix with
glueballs, and their  spectrum  has many interpretations made by different
authors. For instance, Palano \cite{Palano}  suggested a scenario, in which
the states $a_0(980)$, $K_0^*(1430)$, $f_0(980)$, and $f_0(1400)$ form a
nonet. The state $f_0(1500)$ is considered as the scalar glueball.
T\"ornqvist et al.~\cite{Tornqvist} looked upon the states $f_0(980)$ and
$f_0(1370)$ as manifestations of the ground and excited $s\bar s$ states;
and the state $f_0(400-1200)$, as the ground $u\bar u$ state.
Van Beveren et al.~\cite{Beveren} considered
the states  $f_0(400-1200)$ and $f_0(1370)$ as
$u\bar u$ ground states;
and the states $f_0(980)$ and $f_0(1500)$,  as $s\bar s$ ground states.
Two ground states for each $q\bar q$ system occur due to pole doubling, which
takes place for scalar mesons in their model.
Shakin et al.~\cite{Shakin} obtained from a nonlocal confinement model that
the $f_0(980)$ resonance is the ground $u\bar u$ state, and $f_0(1370)$
is the ground $s\bar s$ state. The state $f_0(1500)$ is considered as a radial
excitation of $f_0(980)$. They believed the mass of the scalar glueball to be
1770 MeV.

In our recent papers \cite{ECHAJA1999},
following the methods given in Refs.~\cite{Volk_86,Volk_82,Weiss,YaF},
we showed that all experimentally observed scalar meson states with
masses in the interval from 0.4 to 1.71 GeV can be interpreted
as members of two scalar meson nonets ---
the  ground state of the meson-nonet  and its first
radial excitation. We considered all scalar mesons as $q{\bar q}$
states and took into account the singlet-octet mixing  caused by
the 't Hooft interaction. In \cite{ECHAJA1999}, we obtained
a scalar isoscalar state with mass 1600 MeV and had to choose,
to which of the experimentally observed
states $f_0(1500)$ and $f_0(1710)$ we should ascribe it.
From our analysis of the decay rates calculated in our
model, we found that $f_0(1710)$
better fits to the nonet of quarkonia than $f_0(1500)$.
Therefore, we supposed that the state $f_0(1500)$ contained
a significant component of the scalar
glueball (see \cite{Naris_98,Aniso_98}). However,
the final decision should be made after including the scalar
glueball into the model, and taking  account of its mixing
with quarkonia, which will shift scalar meson masses.

At present, there exist two candidates for the
glueball: $f_0(1500)$ and $f_0(1710)$ \cite{Aniso_98,Tornqvist,Jami_96}.
To describe the properties and  mixing of the glueball with the other
scalar states, one should introduce
an additional scalar isoscalar dilaton field $\chi$ into our model,
in addition to the quarkonia which
have  already been described~\cite{ECHAJA1999}.
For this purpose, one can make use of the idea of approximate
scale invariance of effective Lagrangians based on the dilaton
model.
Such models were studied by many authors
(see, e.~g., \cite{gb:models6,Jami_96,Kusa_93,Andr_86,Elli_84}).
Unfortunately, there is no unique way to introduce
the dilaton field into a chiral  Lagrangian.
This justifies the large number of models dealing with
glueballs.

The guideline, one should follow when introducing
the dilaton field into an effective meson Lagrangian, is
to reproduce  the Ward identity connected
with the scale anomaly. The latter leads to the following equation
for the vacuum expectation value of the divergence of the
dilatation current
 \beq
 \label{Ward}
 \langle\partial_{\mu}S^{\mu}\rangle=\Cg-\sum_{q=u,d,s}m_{q}^0
 \langle\bar qq\rangle,
 \eeq
 \beq
 \Cg=\left({11\over 24}N_c - {1\over 12}N_f \right)
 \left\langle{\alpha\over \pi}G^2_{\mu\nu}\right\rangle, \label{gluoncon}
 \eeq
where $N_c$ is the number of colours; $N_f$,
the number of flavours;  $\langle {\alpha\over\pi}G_{\mu\nu}^2\rangle$ and
$\langle\bar qq\rangle$,
the gluon and quark condensates; $m^0_{q}$, the current quark mass.

In this paper, we are going to use the most natural method of
introducing the dilaton field into the effective Lagrangian
by requiring that, in the chiral limit, our Lagrangian should
be scale-invariant except for the dilaton potential.  To
realize this program, one should multiply all dimensional
parameters of the original Lagrangian (without dilaton) by
a corresponding power of
the dilaton field divided by its vacuum expectation value
$\chi_c$ to preserve the dimensions of model parameters.
Thus, instead of the four-quark
coupling constant $G$, the 't Hooft coupling constant $K$,
ultraviolet cutoff $\Lambda$ (necessary for the
regularization of divergent integrals coming from quark
loops), and the constituent quark masses $m_q$ $(q=u,s)$, one
should use $G(\chi_c/\chi)^2$, $K(\chi_c/\chi)^5$,
$\Lambda(\chi/\chi_c)$ and $m_q(\chi/\chi_c)$.

Current quark masses $m^0_q$ are not multiplied by the
dilaton field and violate scale invariance explicitly, as it
takes place in QCD.  Their contribution to the divergence of
the dilatation current is determined by quark condensates
and disappears in the chiral limit (see (\ref{Ward})).

Omitting, for a moment, the 't Hooft interaction
in our approach (which, to an extent,
is in the spirit of papers \cite{gb:models6,Jami_96,Weiss} ),
we require that
the Lagrangian is scale-invariant in the chiral limit  both
before and after the spontaneous breaking of chiral symmetry (SBCS),
except for  the dilaton potential.
This property can be obtained by considering
(after bosonization when the effective Lagrangian is expressed
in terms of bosonic scalar and pseudoscalar
fields $\sigma$ and $\phi$) the shift of the scalar meson field
$\sigma$
\beq
\sigma=\sigma'-m\frac{\chi}{\chi_c}, \qquad (m^0=0), \label{sigma:shift}
\eeq
where $\langle\sigma'\rangle_0=0,\quad \langle\sigma\rangle_0=-m$,
guaranteing that the relation (\ref{Ward}) is satisfied
\cite{Ebert88}.
The nonzero vacuum expectation value of $\sigma$ appears as a result
of SBCS, and thus, the constituent quark mass is produced.
In the case of nonvanishing current quark masses,
 (\ref{sigma:shift}) changes by including an additional
(non-scaled) mass term
$m^0$ into the r.h.s.
This change  produces an interaction term $\sim {m^0\over
G}\left({\chi\over\chi_c}\right)^2\sigma'$ in the effective Lagrangian
(\ref{LGr})
which breaks both chiral and scale symmetry  just in the way
required by the quark mass term $m^0\bar qq$ of the QCD Lagrangian.

The structure of the paper is as follows. In Section 2, we derive the
usual $U(3)\times U(3)$-flavour symmetric effective Lagrangian with
the 't Hooft interaction and without dilaton fields.  In Section 3,
the dilaton field is introduced into the effective Lagrangian obtained
in Section 2.  Gap equations are investigated in Section 4.  In
Section 5, we derive mass terms and fix the model parameters. The main
decays of scalar isoscalar mesons are calculated in Section 6. Finally,
in the Conclusion, we discuss the obtained results.

\section{Chiral effective Lagrangian with  't Hooft interaction}

A $U(3)\times U(3)$ chiral Lagrangian with the 't Hooft interaction
was investigated in paper \cite{Cimen_99}.  It consists of three terms
as shown in formula (\ref{Ldet}).  The first term represents the free
quark Lagrangian, the second is composed of four-quark vertices as in
the NJL model, and the last one describes the six-quark 't Hooft
interaction~\cite{Dorokhov92} that is necessary to solve the $U_A(1)$
problem.
\ba
L& =& {\bar q}(i{\hat \partial} - m^0)q + {G\over 2}\sum_{a=0}^8
[({\bar q} {\lambda}_a q)^2 +({\bar q}i{\gamma}_5{\lambda}_a q)^2]
-\nonumber\\ &&- K \left\{ {\det}[{\bar q}(1+\gamma_5)q]+{\det}[{\bar
q}(1-\gamma_5)q]
\right\}.
\label{Ldet}
\ea
Here $G$ and $K$ are coupling constants, $\lambda_a\; (a=1,...,8)$ are
the Gell-Mann matrices $\lambda_0 = {\sqrt{2/ 3}}$~{\bf 1}, with {\bf
1} being the unit matrix; $m^0$ is a current quark mass matrix with
diagonal elements $m^0_u$, $m^0_d$, $m^0_s$ $(m^0_u \approx m^0_d)$.

The standard bosonization procedure
for local quark models consists in replacing
the four-quark vertices by Yukawa couplings of quarks with bosonic
fields which enables one to perform the integration over quark fields.
The final effective bosonic Lagrangian appears
then as a result of the calculation of the quark determinant.
To realize this program, it is necessary,
using the method described in
\cite{Cimen_99,Dorokhov92,Vogl_91,Kleva_92}, to go from
Lagrangian (\ref{Ldet})
to an intermediate Lagrangian which contains only four-quark vertices
\ba
&&L = {\bar q}(i{\hat \partial} -
\overline{m }^0)q + {1\over 2}\sum_{a,b=1}^9[G_{ab}^{(-)} ({\bar q}{\tau}_a
q)({\bar q}{\tau}_b q)\nonumber\\
&&\quad +G_{ab}^{(+)}({\bar q}i{\gamma}_5{\tau}_a
q)({\bar q}i{\gamma}_5{\tau}_b q)], \label{LGus}
\ea
where
\ba &&{\tau}_a={\lambda}_a
~~~ (a=1,...,7),~~~\tau_8 = ({\sqrt 2} \lambda_0 + \lambda_8)/{\sqrt
3},\nonumber\\ &&\tau_9 = (-\lambda_0 + {\sqrt 2}\lambda_8)/{\sqrt 3},
\nonumber \\
&&G_{11}^{(\pm)}=G_{22}^{(\pm)}=G_{33}^{(\pm)}= G \pm
4Km_sI^\Lambda_1(m_s), \nonumber \\
&&G_{44}^{(\pm)}=G_{55}^{(\pm)}=G_{66}^{(\pm)}=G_{77}^{(\pm)}= G \pm
4Km_uI^\Lambda_1(m_u), \nonumber \\ &&G_{88}^{(\pm)}= G \mp
4Km_sI^\Lambda_1(m_s),\quad G_{99}^{(\pm)}= G,\nonumber\\
&&G_{89}^{(\pm)}=G_{98}^{(\pm)}= \pm 4{\sqrt
2}Km_uI^\Lambda_1(m_u),\nonumber\\ &&G_{ab}^{(\pm)}=0\quad (a\not=b; \quad
a,b=1,\dots,7),
\label{DefG}
\ea
\ba
    \overline{m}^0_u&=&m^0_u- 32 K m_u m_s
    I^{\Lambda}_1(m_u)I^{\Lambda}_1(m_s) \label{twoloopcorrect1},\\
    \overline{m}^0_s&=&m^0_s- 32 K m_u^2
    I^{\Lambda}_1(m_u)^2\label{twoloopcorrect2}.
\ea
Here $m_u$ and $m_s$ are constituent quark masses and the integrals
\ba
I^{\Lambda}_n(m_a)\!=\!\!{N_c\over (2\pi)^4}\!\!\int\!\! d^4_e k {\theta (\Lambda^2
-k^2)
\over (k^2 + m^2_a)^n}, \; (n\!=\!1,2;\; a\!=\!u,s),
\label{DefI}
\ea
are calculated in the Euclidean metric and regularized by a simple
$O(4)$-symmetric ultraviolet cutoff $\Lambda$. For
$I^{\Lambda}_1(m_a)$ one gets
\beq
  I^{\Lambda}_1(m_a)=\frac{N_c}{16\pi^2}
\left(\Lambda^2-m_a^2\ln\left({\Lambda^2\over m_a^2}+1\right)\right),
\label{I1}
\eeq
where $m_a$ represents a corresponding constituent quark mass%
\footnote{ The notation ``constituent'' quark mass refers here to the
total quark mass appearing in the full quark propagator.
}: $m_u$ or $m_s$.
Note that we have introduced the notation of
constituent quark mass  already here,
although they will be consistently considered only later,
when discussing mass gap equations (compare (\ref{gap:eq:u})
and (\ref{gap:eq:s})) and the
related shift of scalar meson
fields
(see (\ref{sigma_shift}) and (\ref{Lbar})).
 However, as we want to use an effective four-fermion
interaction instead of the original six-quark one, we have to
calculate quark loop corrections for the constant $G$ (see
(\ref{DefG})) using full quark propagators with constituent quark
masses.

In addition to the one-loop corrections to the constant $G$ at
four-quark vertices, we have to  modify the current quark masses $m_a^0$
(see (\ref{twoloopcorrect1}) and (\ref{twoloopcorrect2})).  This is to
avoid the problem of double counting of the 't Hoot contribution in
gap equations which was encountered by the author in
\cite{Kleva_92}. After the redefinition of the constant $G$ and of the
current quark masses, we can guarantee that in the large-$N_c$ limit
the mass spectrum of mesons and the gap equations, derived from the
new Lagrangian with modified four-quark vertices, are the same as those which
are obtained from the original Lagrangian with six-quark vertices.

Now we can bosonize  Lagrangian (\ref{LGus}).
By introducing auxiliary scalar $\sigma$ and pseudoscalar $\phi$
fields, we obtain
\cite{Volk_86,Volk_82,Cimen_99}
\ba
{\mathcal L}(\sigma,\phi) =
-\frac12\sum_{a,b=1}^9\left(
\sigma_a(G^{(-)})^{-1}_{ab}\sigma_b + \phi_a(G^{(+)})^{-1}_{ab}\phi_b
\right)  \nonumber \\
\quad-i~{\rm Tr}\ln \left\{ i{\hat \partial} - \bar m^0+
\sum_{a=1}^9\tau_a(\sigma_a + i\gamma_5 \phi_a) \right\}.
\ea
As we expect, the chiral symmetry is spontaneously broken due
to the strong attraction of quarks in the scalar channel and
the scalar isoscalar fields acquire nonzero vacuum expectation
values $\langle\sigma_a\rangle_0\not=0$ $(a=8,9)$.
These values are related to basic model parameters
$G$, $m^0$ and $\Lambda$ via gap equations as
it will be shown in the next Section.
Therefore, we first have to shift the $\sigma$ fields by  proper values
so that the new fields have zero vacuum expectation values
\beq
\sigma_a=\sigma_a'-\mu_a+\bar\mu_a^0,\qquad \langle\sigma_a'\rangle_0=0,
\label{sigma_shift}
\eeq
where $\mu_a=0, \quad (a=1,\dots ,7)$,
 $\mu_8=m_u$, $\mu_9=-m_s/\sqrt{2}$ and
$\bar\mu_a^0=0, \quad (a=1,\dots ,7)$,
 $\bar\mu^0_8=\bar m_u^0$, $\bar \mu^0_9=-\bar m_s^0/\sqrt{2}$.
After this shift we obtain:
\ba
&&{\mathcal L}(\sigma',\phi) =
\quad L_G(\sigma',\phi)
-i~{\rm Tr}\ln \left\{ i{\hat \partial} -  m\right.\nonumber\\
&&\quad\left. +\sum_{a=1}^9\tau_a(\sigma_a' + i\gamma_5 \phi_a) \right\}.
\label{Lbar}
\ea
where
\ba
&&L_G(\sigma',\phi)=\nonumber\\
&&\quad-\frac12\sum_{a,b=1}^9
(\sigma_a'-\mu_a+\mu_a^0)\left(G^{(-)}\right)^{-1}_{ab}
(\sigma_b'-\mu_a+\mu_a^0)\nonumber\\
&&\quad -\frac12 \sum_{a,b=1}^9
\phi_a\left(G^{(+)}\right)^{-1}_{ab}
\phi_b.  \label{LG}
\ea
From Lagrangian (\ref{Lbar}) we take only those terms  (in momentum space)
which are linear, squared, cubic and quartic
in scalar and pseudoscalar
fields.\footnote{Despite that the scalar fields
are of the main interest in                                       
this paper, we still need  pseudoscalar fields to fix
the model parameters.}
\ba
&& {\mathcal L}(\sigma',\phi)=L_G(\sigma',\phi)+
\tr\Bigl[I_2^{\Lambda}(m)(\sigma'^2+\phi^2)\nonumber\\
&&\quad- 4 m I^\Lambda_1(m)\sigma'+
2I_1^{\Lambda}(m)(\sigma'^2+\phi^2)\nonumber\\
&&\quad-4m^2I_2^{\Lambda}(m)\sigma'^2
+4mI^{\Lambda}_2(m)\sigma'(\sigma'^2+\phi^2)^2\nonumber\\
&&\quad-
I^{\Lambda}_2(m)(\sigma'^2+\phi^2)^2+I^{\Lambda}_2(m)[\sigma'-m,\phi]_{-}^2,
\label{Lagr:bosonized}\\
&& \sigma'=\sum_{a=1}^9\sigma_a\tau_a,\qquad \phi=\sum_{a=1}^9\phi_a\tau_a,
\ea
where  ``tr''
means calculating the trace over  $\tau$-matrix expressions
and $[\dots]_-$ stands for a commutator \cite{Volk_86}.
The expression for $I^\Lambda_1(m_a)$  in
Euclidean metric is given in (\ref{I1}).
The integrals $I^{\Lambda}_2(m_a)$ are also calculated
in Euclidean space-time
\beq
I^{\Lambda}_2(m_a)=\frac{N_c}{16\pi^2}
\left(\ln\left({\Lambda^2\over m_a^2}+1\right)-
{\Lambda^2\over \Lambda^2+m_a^2}\right).
\label{I2}
\eeq
Then, we  renormalize the fields in (\ref{Lagr:bosonized}) so that
the kinetic terms of the effective Lagrangian are of conventional
form, and diagonalize the isoscalar sector.
\ba
&& \bar{\mathcal L}(\sigma^r,\phi^r)=
\bar L_G(\sigma^r,\phi^r)\nonumber\\
&&\qquad+
\tr\Bigl[\frac{p^2}{4}(\sigma^{r\; 2}+\phi^{r\;2})-
4mg I_1^{\Lambda}(m)\sigma^r\nonumber\\
&&\qquad+ 2g^2 I^{\Lambda}_1(m)(\sigma^{r\; 2}+
Z\phi^{r\;2})+\frac14[m,\phi^r]_{-}^2 \nonumber\\
&&\qquad - m^2 \sigma^{r\; 2} +
 m g\sigma^{r}(\sigma^{r\; 2}+Z\phi^{r\;2})-
\frac12[m,\phi^{r}]_{-}[\sigma^{r},\phi^{r}]_{-}\nonumber\\
&&\qquad-{g^2\over 4}((\sigma^{r\; 2}+Z\phi^{r\; 2})^2 -
[\sigma^{r},\phi^{r}]_{-}^2
)\Bigr],
\label{Lagr:bosonized:r}\\
&&\qquad \sigma^{r}=\sum_{a=1}^9\sigma_a^{r}\tau_a,
\qquad \phi^r=\sum_{a=1}^9\phi_a^{r}\tau_a.
\ea
For $\bar L_G$ we have:
\ba
&&\bar L_G(\sigma^r,\phi^r)=\nonumber\\
&&\quad -\frac12\sum_{a,b=1}^9
(g_a\sigma^r_a-\mu_a+\bar\mu_a^0)\left(G^{(-)}\right)^{-1}_{ab}
(g_b\sigma^r_b-\mu_b+\bar\mu_b^0)\nonumber\\
&&\quad-
\frac{Z}{2}\sum_{a,b=1}^9
g_a\phi^r_a\left(G^{(+)}\right)^{-1}_{ab}
g_b\phi^r_b.  \label{LG1}
\ea
Here we introduced
Yukawa coupling  constants $g_a$:
 \beq
\sigma'_a=g_a\sigma^r_a,\qquad \phi_a=\sqrt{Z}g_a\phi^r_a, \label{renorm}
\eeq
\ba
&& g_1^2=g_2^2=g_3^2=g_8^2=g_u^2=[4I^\Lambda_2(m_u)]^{-1},\nonumber\\
&&g_4^2=g_5^2=g_6^2=g_7^2=[4I^\Lambda_2(m_u,m_s)]^{-1}, \nonumber \\
&&\quad g_9^2=g_s^2=[4I^\Lambda_2(m_s)]^{-1}, \\
&&I^\Lambda_2(m_u,m_s)=
{N_c\over (2\pi)^4}\int d^4_e k {\theta (\Lambda^2 -k^2)
\over (k^2 + m^2_u)(k^2 + m^2_s)}=\nonumber\\
&&\quad={3\over (4\pi)^2(m_s^2-m_u^2)}\left[m_s^2\ln\left({\Lambda^2
\over m_s^2}+1 \right)\right.\nonumber\\
&&\quad - m_u^2\ln\left.\left({\Lambda^2\over m_u^2}+1
\right) \right],
\label{ga_0}\\
&& Z=\left(1-\frac{6m_u}{M_{A_1}^2}\right)^{-1}\approx 1.44 \label{Z} ,
\ea
where we have taken into account $\pi$-$A_1$-transitions
leading to an additional $Z$ factor, with $M_{A_1}$ being
the mass of axial-vector meson (see \cite{Volk_86}).
The renormalized scalar and pseudoscalar
fields in (\ref{renorm}) are marked with the superscript $r$.

The mass formulae for isovectors and isodublets follow immediately from
(\ref{Lagr:bosonized:r}). One just has to look up for the coefficients at
$\sigma^{r\; 2}$ and $\phi^{r\; 2}$. There are still nondiagonal
terms in (\ref{LG1}) in the isoscalar sector. This problem is solved by
choosing the proper mixing angles both for the scalars and pseudoscalars
(see e.g. \cite{Cimen_99}).
As we are going to introduce the glueball field, the mixing with
scalar isoscalar quarkonia will change the situation. One has to
consider the mixing among three states, which cannot be described
by a single angle.
For simplicity, in our estimations we
resort to a numerical diagonalization procedure, not to the algebraic one.
Concerning the pseudoscalar sector, one can avail oneself with
the results given in \cite{Cimen_99}.
All what concerns dealing with the glueball is discussed in
the next Section.

\section{Nambu--Jona-Lasinio model with dilaton.}

As we have already mentioned before, we introduce
the glueball into our effective Lagrangian, obtained in the
previous Section, as a dilaton.
For this purpose  we use the following principle. Insofar
as the QCD Lagrangian, in the chiral limit, is
scale invariant, we suppose that our effective meson Lagrangian, motivated by
QCD, has also to be  scale invariant
both before and  after SBCS
in the case when the current
quark masses  are equal to zero.
Note that
the scale anomaly of QCD is reproduced by the dilaton potential.
As a result, we come to the following prescription:
the dimensional model parameters $G$, $\Lambda$, and $K$ are
replaced by the following rule $G\to G(\chi_c/\chi)^2$,
$K\to K(\chi_c/\chi)^5$, $\Lambda\to\Lambda (\chi/\chi_c)^2$,
where $\chi$ is the dilaton field with the vacuum expectation
value $\chi_c$.
Moreover, the
constituent quark masses are replaced by the rule
$m_a\to m_a(\chi/\chi_c)$.
Concerning the current quark masses, the are left unscaled.
This leads to the following formula
\ba
&&        \sigma_a=\sigma_a'-(\mu_a-\bar\mu_a^0+\mu_a^0)
{\chi\over\chi_c}+\mu_a^0\nonumber\\
&&\quad       =\sigma_a'-(\mu_a-\bar\mu_a^0){\chi\over\chi_c}-
\mu_a^0{\chi'\over\chi_c}, \label{sigma:shift:chi}
\ea
where $\mu_a^0=0,\quad (a=1,\dots, 7)$,
$\mu_8^0=m_u^0$, $\mu_9^0=-m_s^0/\sqrt{2}$, and
the definition of $\bar\mu^0$ is given after (\ref{sigma_shift}).
The difference $\bar\mu^0_a-\mu^0_a$ is proportional to
the 't Hooft interaction constant $K$ (see (\ref{twoloopcorrect1}) and
(\ref{twoloopcorrect2})) and
has conventional scale behaviour, therefore it should be scaled in
the same way as $\mu_a$. Note that in the r.h.s of (\ref{sigma:shift:chi})
$\chi'$ denotes the quantum fluctuations of the dilaton field around
its vacuum expectation value $\chi_c$.

Finally, we come to the following Lagrangian:
\ba
&&\bar{\mathcal L}(\sigma^r,\phi^r,\chi)={\mathcal L}(\chi)+
L_{kin}(\sigma^r,\phi^r)+
\bar L_G(\sigma^r,\phi^r,\chi)\nonumber\\
&&\quad+\tr\Bigl[-4m g I^\Lambda_1(m)\sigma^r \left({\chi\over\chi_c}\right)^3\nonumber\\
&&\quad+ 2g^2I^\Lambda_1(m)(\sigma^{r\; 2}+\phi^{r\;2})
\left( {\chi\over\chi_c}\right)^2-
m^2g^2\sigma^{r\; 2} \left({\chi\over\chi_c}\right)^2 \nonumber\\
&&\quad+m g {\chi\over\chi_c}\sigma^r(\sigma^{r\; 2}+\phi^{r\; 2})-
{g^2\over 4}(\sigma^{r\; 2}+\phi^{r\; 2})^2\Bigr].
\label{Lagr:bosonized:chi}
\ea
Here ${\mathcal L}(\chi)$ is the pure dilaton Lagrangian
\beq
{\mathcal L}(\chi)=\frac12(\partial_\nu\chi)^2-V(\chi)
\eeq
with the  potential
\ba
V({\chi})=B\left({\chi\over {\chi}_0} \right)^4\left[ \ln \left({\chi\over
{\chi}_0} \right)^4 -1 \right] \label{chi}
\ea
that has a
minimum at $\chi = \chi_0$, and the parameter $B$ represents the vacuum
energy, when there are no quarks. The curvature of the potential at its
minimum determines the bare glueball mass
\ba m_g = {4\sqrt{B}\over {\chi}_0}.
\label{Defm_g}
\ea
The part $L_{kin}(\sigma^r,\phi^r)$ of Lagrangian (\ref{Lagr:bosonized:chi})
contains pure kinetic terms
\beq
L_{kin}(\sigma^r,\phi^r)=\frac12(\partial_\nu\sigma_a^r)^2+
\frac12(\partial_\nu\phi_a^r)^2
\eeq
to which we  pay no further attention.
The next term reads
\ba
&&\bar L_G(\sigma^r,\phi^r,\chi)=\nonumber\\
&&\;-\frac12\left({\chi\over\chi_c}\right)^2\!\!\!\sum_{a,b=1}^9\!\!\!
\left(g_a\sigma_a^r\!-\!(\mu_a-\bar\mu_a^0+\mu_a^0){\chi\over\chi_c}+
\mu^0_a\right)\left(G^{(-)}\right)^{-1}_{ab}\nonumber\\
&&\;\times
        \left(g_b\sigma_b^r-(\mu_b-\bar\mu_b^0+\mu_b^0){\chi\over\chi_c}
+\mu^0_b\right)\nonumber\\
&&\; -\frac{Z}{2}\left({\chi\over\chi_c}\right)^2\!\!\!\sum_{a,b=1}^9\!\!\!
g_a\phi_a^r\left(G^{(+)}\right)^{-1}_{ab}
        g_b\phi_b^r=\nonumber\\
&&\;-\frac12\left({\chi\over\chi_c}\right)^2\!\!\!\sum_{a,b=1}^9\!\!\!
\left(g_a\sigma_a^r\!-\!(\mu_a-\bar\mu_a^0){\chi\over\chi_c}-
\mu^0_a\frac{\chi'}{\chi_c}\right)\left(G^{(-)}\right)^{-1}_{ab}\nonumber\\
&&\;\times
        \left(g_b\sigma_b^r-(\mu_b-\bar\mu_b^0){\chi\over\chi_c}
-\mu^0_b\frac{\chi'}{\chi_c}\right)\nonumber\\
&&\; -\frac{Z}{2}\left({\chi\over\chi_c}\right)^2\sum_{a,b=1}^9
g_a\phi_a^r\left(G^{(+)}\right)^{-1}_{ab}
        g_b\phi_b^r.
\label{LGr}
\ea
The dilaton field is here expanded around
its vacuum expectation value:
$\chi = {\chi'} + \chi_c$, $\langle\chi\rangle_0 = \chi_c$,  $\langle{\chi'}\rangle_0=0$.

Recall that
the terms proportional to $m^0$ break explicitly chiral and scale
invariance in the same way as the current  mass term of
QCD Lagrangian. Notice also that for our linear $\sigma$-model
(\ref{Lagr:bosonized:chi}), together with the gap
equation (see (\ref{gap:eq:u})), lead to a scale-invariant
pion term \hbox{$\sim M_{\mathbf{\pi}}^2\pi^2(\chi/\chi_c)^2/2$}
instead of the scale-violating term\\ \hbox{$\sim
M_{\pi}^2\pi^2(\chi/\chi_c)/2$}
arising in  nonlinear $\sigma$-models \cite{Elli_84,Ebert88}.

As one can see, expanding (\ref{LGr})
in power series of $\chi$, we can extract a term that is of order $\chi^4$.
It can be absorbed by the term in the pure dilaton potential
which has the same degree of $\chi$.
Obviously, this leads only
to a redefinition of the constants $B$ and $\chi_0$, which anyway
are not known from the very beginning.
Moreover, saying in advance, terms like $\chi^4$ do not contribute to the
divergence of the dilatation current (\ref{Ward}) because of
their scale invariance.

Let us now consider the vacuum expectation value of the
divergence of the dilatation current calculated from the
potential of the effective  meson-dilaton Lagrangian
\ba
&& \langle\partial_{\mu}S^{\mu}\rangle=\left(\sum_{a=8}^9
\sigma_a^r{\partial V\over\partial \sigma_a^r}+
 \chi{\partial V\over\partial \chi}-4V\right)
\Biggr\vert_{\begin{array}{l}\scriptstyle \chi
\quad = \chi_c\, \\[-2mm]
\scriptstyle\sigma_a^r = 0
\end{array}}\nonumber\\
&&\quad =4B\left({\chi_c\over\chi_0}\right)^4-
\sum_{q=u,d,s}m^0_q\langle\bar qq\rangle.
\label{dilaton:current}
\ea
Here $V=V(\chi)+\bar V(\sigma^r,\phi^r,\chi)$,
where $\bar V(\sigma^r,\phi^r,\chi)$
is the potential part of Lagrangian $\bar{\mathcal L}(\sigma^r,\phi^r,\chi)$.
Note that we have simplified (\ref{dilaton:current}), taking into account
that the quark condensates are related to integrals $I_1(m_u)$ and $I_1(m_s)$
as follows
\beq
4m_q I^\Lambda_1(m_q)=-\langle\bar qq\rangle_0,
\qquad (q=u,d,s), \label{I1toQQ}
\eeq
and that these integrals are connected with constants $G^{(-)}_{ab}$
through gap equations, as it will be shown in the next Section.
Comparing the QCD expression (\ref{Ward})
with (\ref{dilaton:current}), one can see that
the quark condensates enter into both formulae in the same way.
Equating the right hand sides of (\ref{Ward}) and (\ref{dilaton:current}),
\beq
\Cg-\sum_{q=u,d,s}m^0_q\langle\bar qq\rangle=4B\left({\chi_c\over\chi_0}\right)^4-
\sum_{q=u,d,s}m^0_q\langle\bar qq\rangle,
\eeq
we obtain the correspondence
\beq
\Cg= 4B\left(\chi_c\over\chi_0\right)^4. \label{Cg}
\eeq
This equation relates the gluon condensate, whose value we take
from other models (see e.g. \cite{Narison96}),
to the model parameter $B$. The next step is to investigate
the gap equations.

\section{Gap equations}

As usual,
gap equations are follow from the requirement that
the terms linear in $\sigma^r$ and $\chi'$ should be absent in our Lagrangian
\ba
&&{{\delta}\bar{\mathcal L}\over {\delta}\sigma_8^r}
\biggr\vert_{(\phi^r,\sigma^r,{\chi'}) = 0}=\!
{{\delta}\bar{\mathcal L}\over {\delta}\sigma_9^r}
\biggr\vert_{(\phi^r,\sigma^r,{\chi'}) = 0}\nonumber\\
&&={{\delta}\bar{\mathcal L}\over {\delta}\chi}
\biggr\vert_{(\phi^r,\sigma^r,{\chi'}) = 0}=0.
\label{var}
\ea
This leads to the following equations
\ba
(m_u-\bar{m}_u^0)(G^{(-)})^{-1}_{88} - {{m_s-\bar{m}_s^0}\over \sqrt2}(G^{(-)})^{-1}_{89}
\quad &&\nonumber\\
-8m_uI^\Lambda_1(m_u) = 0 &&, \label{gapeqbegin} \\
(m_s-\bar{m}_s^0)(G^{(-)})^{-1}_{99} -
{\sqrt2}(m_u-\bar{m}_u^0)(G^{(-)})^{-1}_{98}\quad &&\nonumber\\
-8 m_sI^\Lambda_1(m_s) = 0 && ,\label{gapeq2} \\
4B\left({\chi_c\over {\chi}_0} \right)^3{1\over \chi_0}
\ln \left({\chi_c\over {\chi}_0} \right)^4 + {2A\over \chi_c} = 0.&&
\label{Gapeqs}
\ea
Here
\beq
A={1\over 2}\sum_{a,b=1}^9
(\mu_a-\bar\mu_a^0)(G^{(-)})^{-1}_{ab}\mu_b^0
\label{A}
\eeq
is proportional to the current quark masses $\mu_b^0 \sim m^0_b$
and thereby small.

Using
(\ref{twoloopcorrect1}) and (\ref{twoloopcorrect2}), one can rewrite
the gap equations (\ref{gapeqbegin}) and (\ref{gapeq2})  in
a well-known form \cite{Kleva_92}
\ba
m_u^0&=&m_u-8 G m_u I_1^{\Lambda}(m_u)\nonumber\\
&&-32K m_u m_s I_1^\Lambda(m_u)I_1^\Lambda(m_s),
\label{gap:eq:u}\\
m_s^0&=&m_s-8 G m_s I_1^{\Lambda}(m_s)\nonumber\\
&&-32K(m_u I_1^\Lambda(m_u))^2.
\label{gap:eq:s}
\ea

The equations discussed above allow us to relate the current quark masses
to the rest of model parameters and also to relate the constants $B$
and $\chi_0$ to the gluon condensate and $\chi_c$.
The constituent quark masses, ultraviolet cutoff, and four-quark coupling
constants will be fixed, as usual in NJL, by means of the
Goldberger-Treimann
relation, the $\rho\to\pi\pi$ decay constant, pion weak decay constant
and the mass spectrum of pseudoscalars (For details see
\cite{Cimen_99} and Refs.~therein).
In the next Section we define $\chi_c$, using the bare glueball mass
(without  mixing effects) as a  parameter.

\section{Mass formulae and numerical estimations.}

The potential part of Lagrangian (\ref{Lagr:bosonized:chi})
which is quadratic in fields $\sigma^r$ and $\chi'$
and which we denote as $L^{(2)}$ has the form
\ba
&&L^{(2)}(\sigma^r,\phi^r,\chi') =\nonumber\\
&&\quad -{1\over 2}g_8^2\{[(G^{(-)})^{-1}_{88}
-8I^\Lambda_1(m_u)] + 4m^2_u\}{\sigma^r}^2_8  \nonumber \\
&&\quad-{1\over 2}g_9^2\{[(G^{(-)})^{-1}_{99} -8I^\Lambda_1(m_s)] + 4m^2_s\}{\sigma^r}^2_9
  \nonumber \\
&&\quad-g_8g_9(G^{(-)})^{-1}_{89}{\sigma^r}_8{\sigma^r}_9
-2\left(\Cg-\frac{A}{4}\right)\left(\frac{\chi'}{\chi_c}\right)^2\nonumber\\
&&\quad-\sum_{a,b=8,9}\frac{\mu_a^0}{\chi_c}(G^{(-)})^{-1}_{ab}
g_b\sigma^r_b\chi' \label{BAchi}.
\ea
The dilaton and its interaction with quarkonia does not change the model parameters
$m_u$, $m_s$, $\Lambda$, $G$, and $K$ fixed in our earlier paper
\cite{Cimen_99}
\ba
&&m_u=280\;\mbox{MeV},\;m_s=420\;\mbox{MeV},\;\Lambda =1.25\;\mbox{
GeV},\nonumber\\
&&G=4.38\;\mbox{GeV}^{-2},\;K=11.2\;\mbox{GeV}^{-5}. \label{paramet}
\ea
As it has been already mentioned, after the dilaton field is
introduced into our model, there appear
three new parameters: $\chi_0$, $\chi_c$, and $B$.
To determine these parameters, we use the two equations
(\ref{Cg}) and (\ref{Gapeqs}) and the bare (without mixing effects)
glueball mass
\beq
m_g^2= \frac{4\Cg-A}{\chi_c^2}.
\eeq
We adjust it
so that, in the output, the mass of the heaviest meson would be
1500 MeV  or 1710 MeV, and thereby fix $\chi_c$.
For the gluon condensate, we use the value $(390\; \mbox{MeV})^4$ \cite{Narison96}.
The result of our fit is presented in Table~\ref{T:spectr}
where we show the spectrum of three physical scalar isoscalar states $\sigma_I$,
$\sigma_{II}$ and $\sigma_{III}$. The last one is associated with the glueball.
\begin{table}
\centering
\caption{The masses of physical the scalar meson states $\sigma_I$,
$\sigma_{II}$, $\sigma_{III}$ and the values of  the parameters $\chi_c$,
$\chi_0$, bag constant $B$, and (bare) glueball mass $m_g$ (in MeV)
for two cases: 1) $M_{\sigma_{III}}=1500$ MeV and
2) $M_{\sigma_{III}}=1710$ MeV.}
\label{T:spectr}
\begin{tabular}{||c|c|c|c|c|c|c|c||}
    \hline
    & $\sigma_I$ & $\sigma_{II}$ & $\sigma_{III}$ & $\chi_c$ & $\chi_0$ &$B, [\mbox{GeV}^4]$& $m_g$\\ \hline
  I & 555 & 1075 & 1500 & 191 & 192 & 0.005 & 1480\\
  II & 555 & 1080 & 1710 & 167 & 168 & 0.005 & 1695\\ \hline
\end{tabular}
\end{table}
The parameters $\chi_0$ and $B$ are fixed by the gluon condensate and
constituent quark masses
\beq
\chi_0=\chi_c \exp \left({A \over 2\Cg}\right),
\eeq
\beq
B=\frac{\Cg}{4}\exp \left(-{2A \over \Cg}\right).
\eeq

The mixing of scalar isoscalar fields is described by the matrix $b$ that
connects the nondiagonalized fields $\sigma^r=(\sigma^r_8,\sigma^r_9,\chi')$
with the physical ones $\sigma_{ph}=(\sigma_{I},\sigma_{II},\sigma_{III})$
\beq
 \sigma^r=b\; \sigma_{ph}.
\eeq
The matrix elements of $b$ are given in Table~\ref{mix}.
\begin{table}
\caption{Elements of the matrix $b$,
describing mixing in the scalar isoscalar sector.
The upper table refers to the case  $\sigma_{III}\equiv f_0(1500)$,
the lower one to the case  $\sigma_{III}\equiv f_0(1710)$ }
\label{mix}
\centering
\begin{tabular}{|c|ccc|}
\hline
 & $\sigma_I$ & $\sigma_{II}$ & $\sigma_{III}$\\
\hline
$\sigma_u^r$ & 0.9804 & 0.1865 & $-0.0636$ \\
$\sigma_s^r$ & $-0.1963$ & 0.9535 & $-0.2288$ \\
$\chi'$ & 0.0180 & 0.2368 &  0.9714 \\
\hline
\end{tabular}
\begin{tabular}{|c|ccc|}
\hline
 & $\sigma_I$ & $\sigma_{II}$ & $\sigma_{III}$\\
\hline
$\sigma_u^r$ & 0.9804 & 0.1912 & $-0.0474$ \\
$\sigma_s^r$ & $-0.1965$ & 0.9672 & $-0.1609$ \\
$\chi'$ & 0.0151 & 0.1671 &  0.9858 \\
\hline
\end{tabular}
\end{table}

\section{Decay widths}

Once all parameters are fixed, we can  estimate
the decay widths for the main strong decay
modes of scalar mesons: $\sigma_l\to\pi\pi$, $\sigma_l\to KK$,
$\sigma_l\to \eta\eta$, $\sigma_l\to\eta\eta'$, and  $\sigma_l\to 4\pi$
where $l=I,II,III$.

The amplitudes that describe the decays are relatively simple in our
model. The decays of quarkonia were considered in \cite{Cimen_99}.
Here we only give numerical estimates for their decay widths, where
the mixing with glueball is taken into account (see Table~\ref{mix}).
Below, we discuss only those amplitudes that describe glueball decays.
 The process
$\sigma_{III}\to\pi\pi$ is given by the amplitude
\beq
A_{\sigma_{III}\to\pi\pi}=A_{\sigma_{III}\to\pi\pi}^{g}+
A_{\sigma_{III}\to\pi\pi}^{q}
\eeq
which has been
divided into two parts. The first part represents the contribution
from the pure glueball. It is proportional to the square of the pion mass
\beq
A_{\sigma_{III}\to\pi\pi}^{g}=-\frac{M_\pi^2}{\chi_c}b_{\chi\sigma_{III}}.
\eeq
where $b_{\chi\sigma_{III}}$ represents a corresponding element of the
$3\times 3$ mixing matrix for scalar isoscalar states (see Table~\ref{mix}).
This contribution is small (since it is proportional to the current quark mass
$m^0_u$), and the process is determined by
the second part that describes the decay of the quark component of
the glueball
\beq
A_{\sigma_{III}\to\pi\pi}^{q}= 2 g_u m_u Z b_{\sigma_u\sigma_{III}}.
\eeq
Despite the smallness of mixing, $|b_{\sigma_u\sigma_{III}}|\ll 1$,
this term prevails over the pure glueball contribution because
$M_{\pi}^2$ is noticeably less than $2g_u m_u Z \chi_c  b_{\sigma_u\sigma_{III}}$.
As a result, the decay width of $\sigma_{III}$,
if  is
\beq
\Gamma_{\sigma_{III}\to\pi\pi}=4\; \mbox{MeV}
\eeq
for $\sigma_{III}\equiv f_0(1500)$, and
\beq
\Gamma_{\sigma_{III}\to\pi\pi}=3\; \mbox{MeV}
\eeq
for $\sigma_{III}\equiv f_0(1710)$.  
As one can see this process occurs with a relatively low rate.

In the case of $K\bar K$ channels, the contribution
of the pure glueball is also proportional to the
mass square of the secondary particle, kaon in this case.
But it is rather large, compared to the pion case as $m^0_s\gg m^0_u$.
In the same way, the amplitude can be split into two contributions
\beq
A_{\sigma_{III}\to K\bar K}=A_{\sigma_{III}\to K\bar K}^{g}+
A_{\sigma_{III}\to K\bar K}^{q}
\eeq
where the pure glueball decay into $K\bar K$ is represented by amplitude
\beq
A_{\sigma_{III}\to K\bar K}^{g}= -\frac{2 M_K^2}{\chi_c}b_{\chi\sigma_{III}}.
\eeq
Its value is large and comparable with the quark component contribution
\beq
A_{\sigma_{III}\to K\bar K}^{q}= 2g_u m_u Zb_{\sigma_u\sigma_{III}}-
2\sqrt{2} g_s m_s Z b_{\sigma_s\sigma_{III}}.
\eeq
 In this case, the contribution from the quark component
is provided by both $u(d)$ and $s$ quarks.
In the case that $\sigma_{III}$ is $f_0(1500)$, we have
\beq
 \Gamma_{\sigma_{III}\to K\bar K}=42\; \mbox{MeV},
\eeq
and in the other case ($\sigma_{III}\equiv f_0(1710)$)
\beq
 \Gamma_{\sigma_{III}\to K\bar K}=90\;
\mbox{MeV}.
\eeq
Strange quarks contribute
more and interfere with the pure glueball part, essentially reducing the
decay width (by a factor 3).

The amplitude of the decay of a glueball into $\eta\eta$ and $\eta\eta'$ can also be
considered in the same manner. The only complication is the singlet-octet
mixing in the pseudoscalar sector. The corresponding amplitudes are
\beq
A_{\sigma_{III}\to\eta\eta}=A_{\sigma_{III}\to\eta\eta}^g+A_{\sigma_{III}\to\eta\eta}^q
\eeq
\ba
A_{\sigma_{III}\to\eta\eta}^g&=&-\frac{M_\eta^2}{\chi_c}b_{\chi\sigma_{III}},\\
A_{\sigma_{III}\to\eta\eta}^q&=&2 g_u m_u Z b_{\sigma_u
\sigma_{III}}\sin^2\bar\theta\nonumber\\ &&-
2\sqrt{2} g_s m_s Z b_{\sigma_s\sigma_{III}}\cos^2\bar\theta,
\ea
where $\bar\theta=\theta-\theta_0$, with $\theta$ being the singlet-octet mixing angle
in the pseudoscalar channel, $\theta\approx -19^\circ$ \cite{Cimen_99}, and $\theta_0$ the
ideal mixing angle $\tan\theta_0=1/\sqrt{2}$.
The decay widths thereby are:
\beq
 \Gamma_{\sigma_{III}\to \eta\eta}=25\; \mbox{MeV}
\eeq
for $\sigma_{III}\equiv f_0(1500)$, and
\beq
\Gamma_{\sigma_{III}\to \eta\eta}=42\; \mbox{MeV}
\eeq
for $\sigma_{III}\equiv f_0(1710)$.

For the decay of the glueball into $\eta\eta'$,
we have the following amplitude
\beq
A_{\sigma_{III}\to\eta\eta'}=A_{\sigma_{III}\to\eta\eta'}^g+A_{\sigma_{III}\to\eta\eta'}^q,
\eeq
\ba
A_{\sigma_{III}\to\eta\eta'}^g&=&0,\\
A_{\sigma_{III}\to\eta\eta'}^q&=&-2Z \sin2\bar\theta  (g_u m_u b_{\sigma_u
\sigma_{III}}\nonumber\\
&&+\sqrt{2} g_s m_s b_{\sigma_s\sigma_{III}}).
\ea
The amplitude $A_{\sigma_{III}\to\eta\eta'}^g$ is equal to zero because
there is no decay of a bare glueball into $\eta\eta'$. This process occurs
only due to the mixing between the glueball and scalar isoscalar quarkonia.
The decay widths are as follows
\beq
\Gamma_{\sigma_{III}\to \eta\eta'}=5 \; \mbox{MeV},
\eeq
for $\sigma_{III}\equiv f_0(1500)$,
\beq
 \Gamma_{\sigma_{III}\to \eta\eta'}=5\; \mbox{MeV}.
\eeq
for $\sigma_{III}\equiv f_0(1710)$. The estimate for the decay  $f_0(1500)$ into
$\eta\eta'$ is just qualitative because the decay is allowed only due to
the finite width of the resonance as its mass lies a little bit below the
$\eta\eta'$ threshold. The calculation is made for the mass of $f_0(1500)$ plus
its half-width. For $f_0(1710)$, we have a more reliable estimation since
the mass is large enough for the decay to be possible. One can see that the
order of magnitude for this decay is about 5 MeV. The estimate for
$f_0(1500)$ is not in contradiction with it.

The decays into four pions are estimated as decays proceeding throug
two channels: one with two intermediate scalar resonances ($\chi\to\sigma\sigma\to 4\pi$) and
one with only one intermediate scalar resonance ($\chi\to\sigma 2\pi\to 4\pi$).
Here we neglect the mixing of glueball with quarkonia since the mixing effect
is small (it is  proportional to $m^0_u$).\footnote{
Only the lowest scalar isoscalar resonance is taken into account here.
The contribution from  $f_0(980)$ should be noticeably smaller
because of a large mass and a narrow width of $f_0(980)$.
}
The amplitude describing the decay into $2\pi^+2\pi^-$ is as follows
\beq
A_{\sigma_{III}\to 2\pi^+2\pi^-}=A_{\sigma_{III}\to\sigma\sigma\to 2\pi^+2\pi^-}+
A_{\sigma_{III}\to\sigma 2\pi\to 2\pi^+2\pi^-},
\eeq
\ba
A_{\sigma_{III}\to\sigma\sigma\to 2\pi^+2\pi^-}&=&
-\frac{32m_u^4 Z
M_\sigma^2}{F_\pi^2\chi_c}(\Delta(s_{12})\Delta(s_{34})\nonumber\\
&&+
\Delta(s_{14})\Delta(s_{23})),\\
A_{\sigma_{III}\to\sigma2\pi\to 2\pi^+2\pi^-}&=&
-\frac{16m_u^4 Z}{F_\pi^2\chi_c}(\Delta(s_{12})+\Delta(s_{34})\nonumber\\
&&+\Delta(s_{14})+\Delta(s_{23})),
\ea
where $F_\pi=93$ MeV is the pion week decay constant, $M_\sigma$ is the
mass of the state $\sigma_{I}$. The function $\Delta(s)$ appears due to
the resonant structure of the processes
\beq
\Delta(s)=(s-M_\sigma^2+i M_\sigma \Gamma_\sigma)^{-1},
\eeq
where $\Gamma_\sigma$ is the decay width of the $\sigma_{I}$ resonance (see below).
This function depends on an invariant mass squared $s_{ij}$ defined as follows
\beq
s_{ij}=(k_i+k_j)^2, \qquad (i,j=1,\dots, 4).
\eeq
Here $i$ and $j$ enumerate the momenta $k_i$ of pions $\pi^+(k_1)$, $\pi^-(k_2)$,
$\pi^+(k_3)$, and $\pi^-(k_4)$.
The amplitude describing the decay into $2\pi^0\pi^+\pi^-$ has the form
\ba
&&A_{\sigma_{III}\to 2\pi^0\pi^+\pi^-}=\nonumber\\
&&\quad A_{\sigma_{III}\to\sigma\sigma\to 2\pi^0\pi^+\pi^-}+
A_{\sigma_{III}\to\sigma 2\pi\to 2\pi^0\pi^+\pi^-},
\ea
\ba
A_{\sigma_{III}\to\sigma\sigma\to 2\pi^0\pi^+\pi^-}&=&
-\frac{16m_u^4 Z M_\sigma^2}{F_\pi^2\chi_c}\Delta(s_{12})\Delta(s_{34}),\\
A_{\sigma_{III}\to\sigma2\pi\to 2\pi^0\pi^+\pi^-}&=&
-\frac{16m_u^4 Z}{F_\pi^2\chi_c}(\Delta(s_{12})+\Delta(s_{34})).
\ea
Here $k_1$ and $k_2$ are momenta of the two $\pi^0$, and $s_{12}$ is
their invariant mass squared. The indices 3 and 4 stand for $\pi^+$
and $\pi^-$, respectively.

In the case of the decay into  $4\pi^0$, we have
\ba
A_{\sigma_{III}\to 4\pi^0}&=&A_{\sigma_{III}\to\sigma\sigma\to 4\pi^0}+
A_{\sigma_{III}\to\sigma 2\pi\to 4\pi^0},\\
A_{\sigma_{III}\to\sigma\sigma\to 4\pi^0}&=&
-\frac{16m_u^4 Z
M_\sigma^2}{F_\pi^2\chi_c}(\Delta(s_{12})\Delta(s_{34})\nonumber\\
&&+\Delta(s_{13})\Delta(s_{24})+
\Delta(s_{14})\Delta(s_{23})),\\
A_{\sigma_{III}\to\sigma2\pi\to 4\pi^0}&=&
-\frac{16m_u^4
Z}{F_\pi^2\chi_c}(\Delta(s_{12})+\Delta(s_{13})+\Delta(s_{14})\nonumber\\
&&+\Delta(s_{23})+\Delta(s_{24})+\Delta(s_{34})).
\ea

From our estimation it follows that
in the case, where $\sigma_{III}$ is identified with $f_0(1500)$, we have
the total width
\beq
\Gamma_{\sigma_{III}\to4\pi} =30\; \mbox{MeV},
\eeq
and in the other case ($\sigma_{III}\equiv f_0(1710)$)
\beq
\Gamma_{\sigma_{III}\to 4\pi}=60\; \mbox{MeV}.
\eeq

Let us present the decay widths of $\sigma_I$ and $\sigma_{II}$.
The state $\sigma_{I}$ that we identify with $f_0(400-1200)$ decays mostly
into a pair of pions, and this process determines the width of $\sigma_I$:
\beq
\Gamma_{\sigma_I\to\pi\pi}\approx 760\mbox{ MeV}.
\eeq
The state $\sigma_{III}$ does not affect it noticeably, since
the mixing of the glueball with $u\bar u$ is very small.
Therefore, the decay rate for both $\sigma_{III}\equiv f_0(1500)$
and $\sigma_{III}\equiv f_0(1710)$ is approximately the same in magnitude.

The decay of the state $\sigma_{II}$ that we identify with $f_0(980)$ into
pions is  determined by the quark component and is slightly reduced by the
glueball component because of mixing with the $s\bar s$ quarkonium. We
obtain
\beq \Gamma_{\sigma_{II}\to\pi\pi}=17 \mbox{ MeV},
\eeq
if $\sigma_{III}\equiv f_0(1500)$ and
\beq \Gamma_{\sigma_{II}\to\pi\pi}=15
\mbox{ MeV},
\eeq
if $\sigma_{III}\equiv f_0(1710)$. From experiment, we
know that its decay width lies within the interval from 40 MeV to 100 MeV.
Concerning the process $\sigma_{II}\to \pi\pi$, we obtain a decay width
that is lower than the experimental one. Notice that this prediction is
completely based on singlet-octet mixing following from the 't Hooft
interaction \cite{Cimen_99} where dilaton effects do play a minor role. The
decay into $K\bar K$ can also be taken into account. From experiment we learn that
the decay into $K\bar K$ can contribute about 30\% to the total width \cite{PDG}.
Our estimates for decays of the glueball are collected in
Table~\ref{Gdecays}.
\begin{table}
\caption{The partial and total decay widths (in MeV) of the glueball for two cases:
$\sigma_{III}\equiv f_0(1500)$ and $\sigma_{III}\equiv f_0(1710)$, and experimental
values of decay widths of $f_0(1500)$ and $f_0(1710)$ \cite{PDG}.}
\label{Gdecays}
\centering\begin{tabular}{||r|r|r|r|r|r|r|r||}
\hline
& $\Gamma_{\pi\pi}$ & $\Gamma_{K \bar K}$ & $\Gamma_{\eta\eta}$ & $\Gamma_{\eta\eta'}$ &
$\Gamma_{4\pi}$ & $\Gamma_{tot}$ & $\Gamma_{tot}^{exp}$\\
\hline
$f_0(1500)$ & 4 & 42 & 25 & 5 & 30 & 100 & 112 \\
$f_0(1710)$ & 3 & 90 & 42 & 5 & 60 & 200 & 130 \\
\hline
\end{tabular}
\end{table}

\section{Conclusion}
As it was mentioned in the Introduction, the inclusion of
a scalar glueball into the effective meson Lagrangian is  quite an
ambiguous procedure. The goal of our paper is to find
the most physically justified way to do this.
In the  approach presented above, we assume that
(with the exception of the dilaton potential) scale invariance
holds for the effective Lagrangian  before and after SBCS in the chiral limit.
The terms depending on current quark masses break both the chiral and
scale invariance, in accordance with QCD.
This leads to the requirement that we should introduce
the dilaton field into the constituent quark masses while the
current quark masses remain unscaled.

In this version of a scaled NJL model, the terms that describe mixing of
the glueball with quarkonia are also proportional to current quark masses.
The same is true for the amplitudes describing decays of the glueball into
pairs of pseudoscalars. Insofar as the masses of current quarks are
small in comparison with the other model parameters (constituent quark mass,
$\chi_c$, $\Lambda$, and so on), this results in a small mixing of
the glueball with quarkonia, relatively small  rates for decays of the
glueball into $\pi\pi$, and only slightly changes the decay width
of $f_0(980)\to\pi\pi$ calculated before introducing
the glueball \cite{Cimen_99}.
The decay of the glueball into two pions is mostly
determined by its $q\bar q$ admixture despite the small mixing.
The mixing coefficient here, although being small ($\sim-0.06$ if
$\sigma_{III}\equiv f_0(1500)$), is multiplied
by a relatively large constant describing the decay of the $\sigma$-meson
into a pair of pions.

In the case of the $KK$ channel, both the gluonic and quark components
play an important role since the interference
between the gluonic and quark amplitudes
is large.
The relatively small contribution from the $u\bar u$ component
 slightly increases the decay rate of the glueball.
But the contribution from the $s\bar s$ component reduces the contribution
from the pure glueball by factor 3.

The decay into $\eta\eta$ is mostly determined by the glueball component.
The mixing of the glueball with $\bar ss$  reduces the decay rate
but not significantly. The decay into $\eta\eta'$ is less than
into $\eta\eta$ and is allowed only due to the mixing of quarkonia
with the glueball. This process serves as a measure of this mixing.
However, in the case of $f_0(1500)$, it is difficult to give reliable
estimates for its rate because the process occurs near
the threshold.

Decays into 4 pions are represented by two processes. In the first
one, two intermediate scalar resonances are born by the glueball
with their subsequent decay into two pairs of pions. In the second
process, only one intermediate scalar resonance together with a pair
of pions are produced immediately after the decay of the glueball.
Then, the scalar resonance decays into pions. From our calculations
it follows that the second process is dominant and two scalar resonances
are less probable to appear.

The total width of the third scalar isoscalar state is estimated to
be about 100 MeV for $M_{\sigma_{III}}=1500$ MeV and  200 MeV for
 $M_{\sigma_{III}}=1710$ MeV.  If we assume that the
$f_0(1500)$ state is the scalar glueball,
the total decay width derived from our model is
close to the experimental value (112 MeV). Unfortunately, the
detailed data on the branching ratios of $f_0(1500)$ are not reliable and
controversial \cite{PDG}.

In conclusion, we would like to note that, in our model, the width of
the decay of a glueball into two pions is small, because the amplitude
describing this decay, is proportional to the current mass of $u$-quark
($\sim M_\pi^2\sim m^0_u$) and does not depend on momenta. The latter
in the chiral limit formally disagrees  with the low-energy theorems
obtained in paper \cite{Shifman}. In general, we could consider a version of
our model containing momentum-dependent vertices, whose
momentum dependence is in agreement with these low-energy theorems.
However, such a momentum dependence of the amplitude leads to too large
decay width of a heavy glueball (see \cite{Elli_84}), which  contradicts
the experimental data. This witnesses to the fact that these low-energy theorems
are not justified to be applied
in the case of a heavy glueball.

The results obtained here  correspond to the leading order in
$1/N_c$ expansion (Hartree--Fock approximation).
Next-to-leading order corrections can to an extent change the final results.
Note also that, in the energy region under consideration ($\sim 1500$ MeV),
we work on the brim of the validity of exploiting the chiral symmetry
that was used to construct our effective Lagrangian.
Thus, we can consider our results as rather qualitative.
Nevertheless, a satisfactory agreement with experimental data is
obtained for the total width of $f_0(1500)$.

We are going to use this approach in our future work
for  describing both glueballs and ground and radially excited scalar
meson nonets which lie it the energy interval from 0.4 to 1.71 GeV.
Small mixing angles make us hope that introducing the glueball
into our model will not change the whole picture dramatically.

\begin{acknowledgement}
We are grateful to Prof.~S.B.~Gerasimov, Dr.~A.E.~Dorokhov,
and Dr. N.I.~Kochelev  for useful discussions. The work is
supported by RFBR Grant 00-02-17190, the Heisenberg-Landau program 2000
and the Graduiertenkolleg ``Elementarteilchenphysik'' of
the Humboldt University, Berlin.
\end{acknowledgement}

\end{document}